\documentclass{ws-ijmpcs}

\def\be{\begin{eqnarray}}
\def\ee{\end{eqnarray}}

\def\MeV{\mbox{MeV}}

\newcommand{\msun}{\mbox{$M_\odot$}}
\def\lsim{\lesssim}

\newcommand{\nn}{\nonumber}

\begin{document}

\markboth{Kim,Lee and Rho}
{Triple layered compact star with strange quark matter}

%
\catchline{}{}{}{}{}
%

\title{Triple layered compact star with strange quark matter}

\author{Kyungmin Kim}

\address{Department of Physics, Hanyang University, \\
Seoul 133-791, Korea
\\
kmkim82@hanyang.ac.kr}

\author{Hyun Kyu Lee}

\address{Department of Physics, Hanyang University, \\
Seoul 133-791, Korea  and \\
Asia Pacific Center for Theoretical Physics, Pohang 790-784, Korea\\
hyunkyu@hanyang.ac.kr}

\author{Mannque Rho}

\address{Department of Physics, Hanyang University, \\
Seoul 133-791, Korea and \\
Institute de Physique Theorique,  CEA Saclay, 
F-91191 Gif-sur Yvette, France\\
mannque.rho@cea.fr}

\maketitle

\begin{history}
\received{Day Month Year}
\revised{Day Month Year}
\end{history}

\begin{abstract}
We explore the possibility of three phases in the core of neutron star in a form of triple layers. From the center, strange quark matter, kaon condensed nuclear matter and  nuclear matter  form a triple layer.  We discuss how the phase of strange quark matter is smoothly connected to kaon condensed nuclear matter phase. We also demonstrate that the compact star with triple layered structure  can be a model compatible with the 1.97-solar-mass object PSR J1614-2230 recently observed.

\end{abstract}

\ccode{PACS numbers: 97.60.Jd, 26.60.Kp}

\section{Introduction}

The basic question raised by the recent observation of a 1.97-solar-mass ($M_{\odot}$) neutron star, PSR J1614-2230, using Shapiro delay with Green Bank Telescope \cite{2solar} , is  whether the new degree of freedom such as hyperons, meson condensations (pions, kaons, ...), strange quark matter, and  the multitude of other phases are relevant in explaining the existence of high mass compact stars.   The density of the central core is found to be much higher than the normal nuclear density, $n_0$, up to $\sim 10 n_0$.

Up to $n_0$, the role of strange degrees of freedom is not significant, but they must figure at some density above $n_0$ in the form  of strange mesons (kaons) and strange baryons (hyperons). In addressing this issue, we will take a simplified effective field theory approach. Instead of treating both kaons and hyperons at the same time, we will take an effective Lagrangian consisting of kaons coupled to nucleons  in the sense that the hyperons are integrated out. It has been first suggested  by Kaplan and Nelson~\cite{kaplan-nelson} that  the kaons figure in a  condensate state.

We find that   kaon condensation will play a role of a ``doorway" to quark matter in a smooth manner\cite{KLR}.  Hence   we can consider compact stars with three phases,  nuclear matter (NM) in the outer-layer region,  kaon condensed nuclear matter (KNM), and (2) strange quark matter (SQM) in the inner most region.

We explore how the structure with  KNM and SQM in the core region affects the microscopic observables [e.g., density-dependent electron (or kaon) chemical potential] and macroscopic observables (e.g., mass and radius) of the compact stars.

The outline of this paper is as follows.  In Sec. II, the simplified model for kaon condensation that we use is described. We also discuss in detail the microscopic properties of nucleonic matter with kaon condensation and the EoS.  In Sec. III, the stellar structure with SQM driven by kaon condensation obtained by integrating the Tolman-Oppenheimer-Volkov (TOV) equation is  discussed.  Throughout this work, we set $\hbar = c = 1$ and assume zero temperature. Finally, the results are summarized and discussed in Sec. IV.

\section{Kaon Condensation and SQM}

\subsection{S-wave kaon condensation}
We consider the simplest effective Lagrangian relevant for  kaon condensation in the form~\cite{kuniharu}
\be
{\cal L} = {\cal L}_{KN} + {\cal L}_{NN} \label{Leff}\ee
where
\be
{\cal L}_{KN} &=& \partial_{0} K^- \partial^{0} K^+ - m^2_K K^+ K^-
+ \frac{1}{f^2} \Sigma_{KN}(n^{\dagger} n + p^{\dagger}p) K^+ K^-\nonumber \\
&+& \frac{i}{4f^2} (n^{\dagger} n + 2 p^{\dagger}p) (K^+\partial_{0} K^- - K^- \partial_{0} K^+) \label{LKN}\\
{\cal L}_{NN} &=&
   n^{\dagger} i \partial_{0} n +  p^{\dagger} i \partial_{0} p - \frac{1}{2m}( \vec{\nabla} n^{\dagger}\cdot \vec{\nabla} n + \vec{\nabla} p^{\dagger} \cdot \vec{\nabla} p)
   - V_{NN} \label{LNN}
\ee
Here the fourth term in Eq.~(\ref{LKN}) is the well-known Weinberg-Tomozawa (WT) term  with $f$ identified with the pion decay constant $f_\pi$ in the {\em matter-free} space and $\Sigma_{KN}$ is the $KN$ sigma term.
The validity of using such simple lagrangian involves many  sophisticated and subtle issues to be resolved, which has been discussed in the literatures\cite{KLR}. Now  we write the Hamiltonian in the form
\be
{\cal H} = {\cal H}_{KN}  + {\cal H}_{NN},
\ee
where
\be
{\cal H}_{KN} &=&  \partial_{0} K^- \partial^{0} K^+  + [m^2_K  -
\frac{n}{f^2} \Sigma_{KN}] K^+ K^-, \label{hkn}\\
{\cal H}_{NN} &=& \frac{3}{5}E_F^0 \left(\frac{n}{n_0}\right)^{2/3}
n + V(n) + n \left(1-2\frac{n_p}{n}\right)^2 S(n). \nonumber\\
\ee
For $V(n)$,  the potential energy of NM, and $S(\rho)$, the symmetry energy factor, there are numerous forms for them in the literature, all fine-tuned to fit available experimental data~\cite{bal}. We  take one convenient parametrization ($'x=-1'$ in \cite{bal}) referred to as ``LCK'' in this work.

Adopting the frame work developed by Baym\cite{baym}, the amplitude of kaon condensation, $K$, and kaon chemical potential, $\mu_K$, are defined by the ansatz
\be
K^{\pm} = K e^{\pm i \mu_K t}.
\ee
The kaon condensation condition for $K \neq 0$ is obtained by extremizing the classical action,
\be
m_K^2 - \mu_K^2 = \mu_K \frac{n_n + 2n_p}{2f^2} + \frac{n}{f^2}\Sigma_{KN}, \label{kaoncond}
\ee
which can be solved to get $\mu_K$ or equivalently $m_K^*$ in Eq.~(\ref{mko}). $\Sigma_{KN}$  is only one parameter left undetermined in the kaonic sector. We took  it as a parameter  in  a reasonable range, $200 {\rm MeV}\lsim \Sigma_{KN}< 400$ MeV .

The contributions to the energy density and the pressure from kaon condensation are given by
\be
\epsilon_K &=& \left( m_K^2 + \mu_K^2 - \frac{n}{f^2} \Sigma_{KN} \right) K^2, \label{ek} \\
P_K &=& - \left(m_K^2 - \mu_K^2\right) K^2. \label{pk}
\ee
One can see that the kaon condensation gives a negative contribution to the total pressure for $\mu_K < m_K$.  This soften the EoS substantially leading to smaller mass for the compact star with kaon condensation\cite{bb}.

 The energy density and the pressure of the system are given by
\be
\epsilon &=& \tilde{V}(n) + \rho (1- 2 x)^2 S(n) + \epsilon_{lepton}
+ \Theta(K)\epsilon_K, \label{nmknmene}\\
P &=& n^2 \frac{\partial V(n)/n}{\partial n} +
n^2(1-2 x)^2\frac{\partial S(n)}{\partial n} + P_{lepton}
+\Theta(K) P_K, \label{nmknmpre}
\ee
where $x=n_p / n$.

 The underlying physics  for having kaon condensation, which has relatively lager vacuum mass,  is the decrease of the effective mass denoted as $m_K^*$, which  is basically a function of $m_K,n, x$ owing to the kaon-nucleon interactions:
 \be
 m_K^*= \mu_K(m_K,n,x , ...). \label{mko}
 \ee
 The equilibrium for  the weak process, $n \rightarrow p + K^-$,
\be
\mu_n - \mu_p = m_K^*,
\ee
determines the threshold of kaon condensation $n_t$. Beyond the kaon condensation, where $ m_K^*$ can be identified as the kaon chemical potential $\mu_K$, the chemical equilibrium is reached as
\be
\mu_n - \mu_p &=& \mu_e = \mu_{\mu} \equiv \mu, \nn \\
\mu_K &=& \mu,
\ee
where
\be
\mu &=& 4(1 - 2 x)S_N(n_c) + \mu \tilde{F}(K,\mu,n_c),  \label{betaeq3} \\
\mu_K &=& \frac{1}{2}[\frac{n_K}{K^2} - \frac{n(1+x)}{2f^2}].
\ee
The charge neutrality condition gives
\be
n_p =n_e + n_{\mu} +  \Theta(K)n_K\ .  \label{neutral}
\ee
Equations (\ref{mko}), (\ref{betaeq3}) and  (\ref{neutral})  are the basic equations to be solved to calculate the EoS of KNM.

\subsection{SQM Driven by Kaon Condensation}
We now turn to the possibility that kaon condensation can drive the dense system to a SQM at the critical density $n_c$ defined by the condition, $\mu_K =0$.   From $n_t$ up to the critical density for chiral restoration $n_c$, the weak equilibrium condition reads
\be
\mu=4(1 - 2 x)S_N(n_c) + \mu \tilde{F}(K,\mu,n_c).
\ee
As the kaon chemical potential -- equivalently effective mass -- approaches 0 at the critical density,  the solution $x=\frac{n_p}{n}=1/2$  appears naturally at the phase boundary,  provided $\tilde{F}(K,n_c) \neq 1$, which is assumed to be valid for the range of density we are concerned with.

The chemical equilibrium (via confinement-deconfinement) reads
\be
\mu_n -\mu_p &=& \mu_d-\mu_u,\label{chemq} \\
\mu_{K^-} &=& \mu_s - \mu_u \label{chemk}
\ee
at the phase boundary.
Note that the strange quark is required at the boundary, which implies that there should be a SQM for $n > n_c$.
Because $\mu(=\mu_K) =0$ , we have from Eqs.~(\ref{chemq}) and (\ref{chemk})
\be
\mu_u = \mu_d = \mu_s. \label{musqm}
\ee
This is the chemical potential relation for the SQM.  In the masselss quark limit,
it implies that
the three flavors  have the same number densities\footnote{In a more realistic calculation we have
to take into account of strange quark mass and electrons and/or muons: $n_u =n_d \neq n_s$.},
\be
n_u =n_d =n_s =n_Q
\ee
Then the charge neutrality
\be
\frac{2}{3}n_u - \frac{1}{3}n_d - \frac{1}{3}n_s = 0
\ee
is automatically satisfied and there is no  need for additional leptons.
Kaon condensed nuclear matter will naturally go over to the SQM in $SU(3)$ symmetric phase in the massless limit.
In this simple picture, the KNM leads naturally to a SQM.

The EoS of SQM for $n > n_c$ is  given by
\be
\epsilon_{SQM}  &=& a_4\frac{9}{4\pi^2}\mu_q^4 + B ~~ = ~~ 4.83a_4n^{4/3} + B,\\
P_{SQM}        &=& a_4\frac{3}{4\pi^2}\mu_q^4 - B ~~ = ~~ 1.61a_4n^{4/3} - B,
\ee
where $B$ is the bag constant~\cite{mitbag}. {Here $a_4$ denotes the perturbative  QCD correction~\cite{LP-GFEST,lapr}, which takes the value  $a_4 \leq 1$. The equality holds for SQM without QCD corrections.}

The possibility of  the confinement-deconfinement phase transition to occur at $n_c$  depends on whether  the
phase boundary matching condition
\be
P_{KNM}(n_c) =  P_{SQM}(n_c^Q) \label{pmatch}
\ee
is satisfied.  In fact it depends on the parameters used in this work.  If we  take $\Sigma_{KN}\simeq 259$ MeV, the pressure matching can be satisfied at the critical density $n_c= 6.36 n_0$.  The kaon threshold density is also obtained as $n_t = 4.20 n_0$.   Using chemical equilibrium conditions we get the critical quark number densities, $n^Q_c= 19.08 n_0$. One can see that $n^Q_c/3$ are not much different from $n_c$.  Given the critical densities, we can find a possible set of parameters, $a_4$ and $B^{1/4}$. For example, with $a_4 = 0.624$, we get $B^{1/4}= 101.15\MeV$.

\section{Compact star with triple-layered  structure}
\begin{figure}[t!]
\begin{center}
\includegraphics[width=8.6cm]{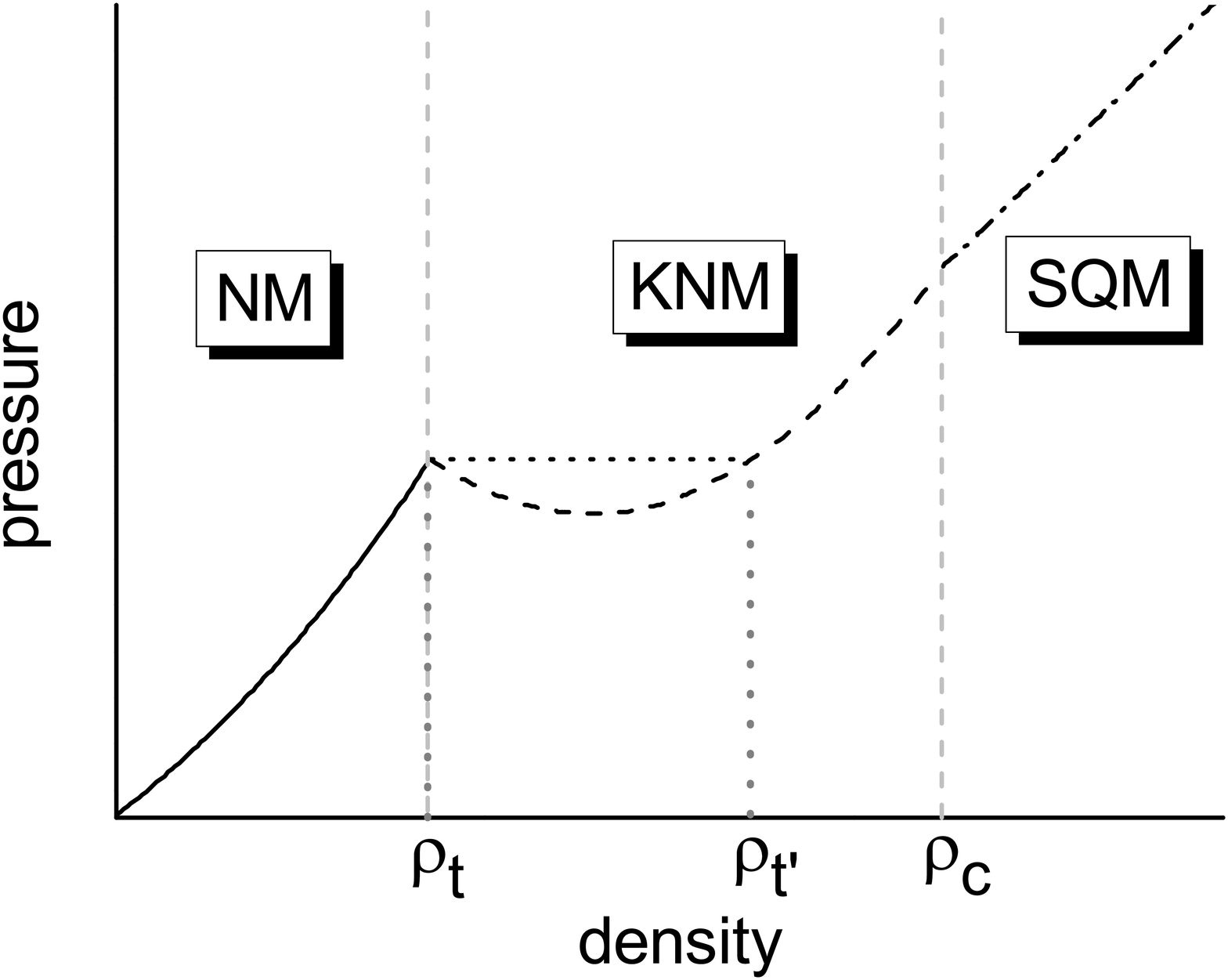}
\caption{The schematic phase diagram for NM-KNM-SQM.}
\label{nkqpre}
\end{center}
\end{figure}

With the parameter set discussed in the previous section,  we can  consider a triple-layered structure shown schematically in Fig.~\ref{nkqpre} consisting of NM, KNM, and SQM from the outer layer to the core part.
In contrast to the case considered above, there will be an unstable region between $n_t$ and $n_{t'}$, for which we have to resort to Maxwell or Gibbs construction. In the Maxwell construction, it turns out to be impossible to match the chemical potential, which signals instability.  In the Gibbs construction, however, there appears a mixed phase of NM and KNM. In this work,  we take a rather simple approach, namely, allow the discontinuity of density and chemical potential by assuming that NM with the density $n_t$  changes into KNM with  the density $n'_t =5.96 n_0$ at the phase boundary defined by
\be
P(n_t) = P(n_{t'}).
\ee

We use Eqs.~(\ref{nmknmene}) and (\ref{nmknmpre}) with Eqs.~(\ref{ek}) and (\ref{pk}) to integrate the TOV equation
\be
\frac{dM}{dr} &=& 4 \pi \epsilon r^2, \nonumber \\
\frac{dP}{dr} &=& - \frac{G M \epsilon}{r^2} \left(1 + \frac{P}{\epsilon}\right) \left(1 + \frac{4 \pi r^3 P}{M}\right)
\left(1 - \frac{2GM}{r}\right)^{-1}, \label{toveqn}
\ee
where $M(r)$ is the mass  enclosed inside the  radius $r$.

\begin{figure}[t!]
\begin{center}
{\includegraphics[width=8.6cm]{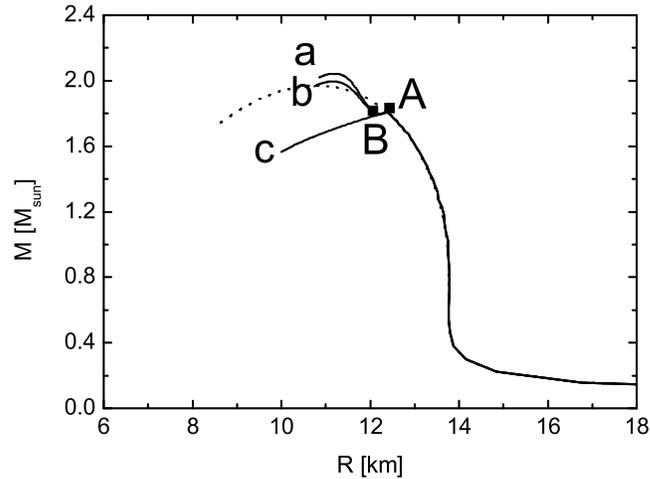}} \hfill
\caption{The M-R sequences for LCK in the text.  The dotted line denotes NM.  The dashed-dotted line between A and B denotes the double-layered (NM-KNM) system.  The solid lines, a, b and c,  denote the triple-layered (NM-KNM-SQM) system with the QCD corrections with $a_4 = 0.59, 0.624$ and $1$, respectively. }
\label{MR-}
\end{center}
\end{figure}

In Fig.~\ref{MR-} the resulting mass-radius relation is plotted.  There is   no sharp jump in density at the phase boundary of  SQM and KNM, which implies that the kaon-driven SQM transition is a rather smooth transition in this scenario. We get the maximum mass, $M_{max} = 1.99\msun$, radius, $ R|_{{}_{M=M_{max}}}= 11.12$km, and central density, $n_{(c)}=11.7n_0$ for the three-layered structure  with the QCD corrections, $a_4 = 0.624$.

\section{Summary and Discussion}

Strange quark matter can appear as a result of confinement-deconfinement phase transition at higher density.  This phase transition is basically a strong interaction process. We discuss the characteristics of the phase change when   kaons are present in the form of condensate. Then there is already non-vanishing strangeness in the NM up to the phase boundary. In this case,  we can imagine the confinement-deconfinement phase transition taking place constrained by the weak equilibrium in the kaon condensed matter, leading to SQM. It leads to the scenario in which kaon condensation joins smoothly a SQM. This scenario was predicted a long time ago by a $K^{-}$ bound to a skyrmion embedded in hypersphere~\cite{forkel}.

For kaon-nuclear interactions, we take the simplest chiral effective Lagrangian at tree order. As for nuclear interactions, we take LCK's empirically parameterized form for the energy density and symmetry energy.  Although there are  potentially serious effects that are missing in this simple  treatments, we take it as a viable basis of our analysis \cite{KLR}.  We illustrate that the compact star with an NM-KNM-SQM structure  is possible with the parameters that are not excluded by theory or phenomenology.

With  $\Sigma_{KN} \simeq 259$ MeV,  the three-layer structure of NM, KNM, and SQM with an appropriate bag constant is possible beyond the critical density, $6.36 n_0$. And the maximum mass of a neutron star with the three-layer structure can be made consistent with the recent observation of PSR J1614-2230.  The EoS obtained is consistent roughly with the analysis by  \"{O}zel et al. \cite{ozel}.  More details on the strangeness distribution and density profile, which might be  relevant for  observational opportunities, for example gravitational wave radiation in compact star merger process, will be discussed elsewhere\cite{KLR2}.

\section*{Acknowledgments}
 This work is supported by WCU (World Class University) program: Hadronic Matter under Extreme Conditions through the National Research Foundation of Korea funded by the Ministry of Education, Science and Technology (R33-2008-000-10087-0). KK is partially supported by the the National Institute for Mathematical Sciences (NIMS) in Korea.  This work is partially based on the discussions during the Second Year of APCTP-WCU Focus Program ``From dense matter to compact stars in QCD and hQCD" at APCTP in Pohang, Aug. 16-25, 2011.

\def\bi{\bibitem}

{}

\end{document}